\documentclass[twocolumn, showpacs, preprintnumbers, amsmath, amssymb, superscriptaddress, pra]{revtex4}
\usepackage{graphicx}
\usepackage{dcolumn}
\usepackage{bm}
\usepackage{epsf}
\usepackage{amssymb}
\DeclareGraphicsExtensions{.eps}
\begin{document}

\author{D.S. Simon}
\affiliation{Dept. of Electrical and Computer Engineering, Boston
University, 8 Saint Mary's St., Boston, MA 02215}
\author{A.V. Sergienko}
\affiliation{Dept. of Electrical and Computer Engineering, Boston
University, 8 Saint Mary's St., Boston, MA 02215}
\affiliation{Dept. of Physics, Boston University, 590 Commonwealth
Ave., Boston, MA 02215}

\begin{abstract}
We investigate cancellation of spatial aberrations
induced by an object placed in a quantum coincidence interferometer
with type-II parametric down conversion as a light source. We
analyze in detail the physical mechanism by which the cancellation
occurs, and show that the aberration cancels only when the object
resides in one particular plane within the apparatus. In addition,
we show that for a special case of the apparatus it is possible to
produce simultaneous cancellation of {\it both} even-order and
odd-order aberrations in this plane.
\end{abstract}

\title{Spatial-Dispersion Cancellation in Quantum Interferometry}


\maketitle

\section{Introduction and background}

\subsection{Introduction}
Aberration or spatial dispersion occurs when light passing through
or reflecting off of an object gains unwanted phase-shifts that
vary in the transverse spatial direction (orthogonal to the
optical axis). These phase shifts are "unwanted" in the sense that
they differ from those obtained from Gaussian optics and cause
distortions of the outgoing wavefronts. Mathematically, we can
represent the aberrations by pure imaginary exponentials $e^{i\phi
({\mathbf x})}$, where ${\mathbf x}$ is the transverse distance.
Often $\phi ({\mathbf x}) $ may be expanded into a power series in
$|{\mathbf x}|$ and separated into even and odd orders,
\begin{eqnarray} \phi ({\mathbf x}) &=&\phi_{even} ({\mathbf x})+\phi_{odd} ({\mathbf
x}),\\ \phi_{even} ({\mathbf x}) &=&\sum_j a_{2j}
r^{2j}P_{2j}(\theta ) ,\\ \phi_{even} ({\mathbf x}) &=&\sum_j
a_{2j+1} r^{2j+1}P_{2j}(\theta ).\end{eqnarray} Here, $r=|{\mathbf
x}|$, while $P_{2j}(\theta )$ and $P_{2j+1}(\theta )$ are
polynomials in $\sin\theta$ and/or $\cos\theta$. Usually, the
expansion is expressed in terms of Seidel or Zernike polynomials
(\cite{bornwolf, buchdahl, wyant}), but for our purposes the
details of the expansion are not important. The important point
here is simply that the even order terms are symmetric under
reflection, $\phi_{even} ({\mathbf x})=\phi_{even} (-{\mathbf
x})$, while the odd terms are antisymmetric, $\phi_{odd} ({\mathbf
x})=-\phi_{odd} (-{\mathbf x})$.

In the papers \cite{bonato1} and \cite{bonato2}, a particular type
of interferometric device was described, and it was shown that if
an object was placed in either arm of this device, then all
even-order phase shifts introduced by the object will cancel in a
temporal correlation experiment. The effect is very similar to the
even-order frequency-dispersion cancellation first described in
\cite{franson} and \cite{steinberg}. As a light source, the
aberration-cancellation experiment used photon pairs produced via
spontaneous parametric downconversion (SPDC). The cancellation
effect depended on the entanglement of the transverse spatial
momenta in the resulting entangled photon pairs.

In this paper we reexamine the setup of \cite{bonato1} and
\cite{bonato2} with two purposes in mind. After reviewing the
apparatus and the even-order aberration cancellation effect in the
next subsection, we first show (in section II) that for a special
case of the apparatus we can in fact cancel {\it all} aberration,
both even-order and odd-order. This cancellation only occurs when
the sample is placed in one particular plane, and opens up the
possibility of cancelling sample-induced abberation in dynamic
light scattering \cite{berne, pecora}, fluorescence correlation
spectroscopy \cite{maiti}, or other temporal correlation-based
experiments. Our second purpose (carried out in section III) is to
analyze in more detail the results for the coincidence rate, in
order to better understand the physical mechanisms involved in
aberration cancellation. In section IV we discuss the conclusions
that can be drawn from these results.

Note that, because we are motivated by the desire to cancel
aberrations, we will use the phrase "aberration-cancellation" for
convenience throughout this paper, but in fact we mean the
cancellation of {\it all} phase shifts arising in a given plane,
not just the subset that differ from the predictions of Gaussian
optics. In other words, "aberration-cancellation" here means that
only the intensity of the light is affected by the object, not the
phase. So, for example, the placement in the object plane of an
ideal lens, whose operation depends on second order phase shifts,
should have no focusing power at this point; it will be as if the
lens is not there.

\subsection{Even-Order Aberration-Cancellation}

Consider the setup shown in figure 1. In the main part of the
apparatus, the two branches each consist of a Fourier transform
system containing lenses of focal length $f$ and a sample
providing a modulation $G_j({\mathbf y})$ of the beam, where
$j=1,2$ labels the branch and ${\mathbf y}$ is the transverse
distance from the optic axis. The $G_j$ represent objects or
samples whose properties we wish to analyze, and the goal is to
cancel optical aberrations introduced by the samples. The case
where there is a sample only in branch 1 is included by simply
setting $G_2=1$, but we will keep the more general two-sample
case; we will see later that the extra generality pays off by
allowing useful additional effects. A controllable time delay
$\tau$ is inserted in one arm of the interferometer. Since we will
be referring to it often, we give a name to the plane containing
the samples, denoting this plane by $\Pi$. The $\Pi$-plane is
simultaneously the back focal plane of the first lens and the
front focal plane of the second. The two lenses together form a 4f
Fourier transform system. We will examine in a later section what
happens when the sample is moved out of the $\Pi$-plane.
Throughout this paper, we assume that the sample is of negligible
thickness compared to all of the other distances involved in the
apparatus. We will refer to the photon in the upper branch (branch
1) as the signal and the photon in branch 2 as the idler. The
polarizing beam splitter sends the horizontally polarized photon
into the upper (signal) branch and the vertically polarized photon
into the lower (idler) branch.

Photons are fed into the system by a continuous wave laser which
pumps a $\chi^{(2)}$ nonlinear crystal, leading to collinear type
II parametric downconversion. The frequencies of the two photons
are $\Omega_0\pm \nu$, while the transverse momenta are $\pm
{\mathbf q}$. For simplicity, assume the frequency bandwidth is
narrow compared to $\Omega_0$. The two photons have total
wavenumbers ${{\Omega_0\pm \nu}\over c}$, which will be
approximated by $k={{\Omega_0}\over c}$ where appropriate. The
downconversion spectrum is given by
\begin{equation}\Phi({\mathbf q},\nu )= sinc\left[ {{L\Delta ({\mathbf q},\nu )}\over 2}\right] e^{i{{L\Delta ({\mathbf q},\nu )}\over 2}}
.\end{equation} Here, $L$ is the thickness of the nonlinear
crystal and for type-II downconversion we have \begin{equation}
\Delta ({\mathbf q},\nu ) =-\nu D +M\hat {\mathbf e_2}\cdot
{\mathbf q}+{{2|{\mathbf q}|^2}\over {k_{pump}}} .\end{equation}
$D$ is the difference between the group velocities of the ordinary
and extraordinary waves in the crystal, and $M$ is the spatial
walk-off in the direction $\hat{{\mathbf e_2}}$ perpendicular to
the interferometer plane. The last term in $\Delta$ is due to
diffraction as the wave propagates through the crystal.

The parametric downconversion process may be described by a
Hamiltonian of the form \begin{equation} \hat H=i\hbar \chi\hat
a_s^\dagger \hat a_i^\dagger +H.C., \end{equation} where $\hat
a_s$ and $\hat a_i$ are annihilation operators for the signal and
idler photons. The constant $\chi $ includes the amplitude of the
classical pump field. Applying the time evolution operator
$e^{-i\hat H t/\hbar}$ to the vacuum state, we find that the
wavefunction entering the apparatus from the crystal can be
written as
\begin{equation}|\Psi(t)\rangle = (1-|\eta |^2/2)|0\rangle
+ \eta|\Psi_{2}\rangle +\eta^2|\Psi_{4}\rangle +\dots ,
\end{equation} where $\eta =\chi t$, and $|\Psi_{2n}\rangle $
represents a term with $n$ photons in the signal mode and $n$ in
the idler mode. For parametric downconversion we operate in the
regime where $|\eta |<<1$, so that terms higher than
$|\Psi_2\rangle$ may be neglected. In addition, the vacuum term
may be ignored since it will not contribute to coincidence
detection. Thus, effectively our wavefunction is given by
\begin{equation}|\Psi\rangle \approx |\Psi_2\rangle=\int dq\; d\nu\; \Phi({\mathbf q},\nu ) \hat
a_s^\dagger ({\mathbf q},\Omega_0+\nu )\hat a_i^\dagger (-{\mathbf
q},\Omega_0-\nu ) |0\rangle .\end{equation}

Note that $G_1$ and $G_2$ could be produced by two separate
objects at two separate points in space, in which case we would
need to use a polarizing beam splitter (PBS) to separate the
incoming beams. Alternatively, $G_1$ and $G_2$ could both be
produced by a single object which acts differently on the two
polarization states, in which case we could dispense with the PBS.

In the detection stage, two bucket detectors $D_1$ and $D_2$ are
connected in coincidence. We add adjustable irises with aperture
functions $p_1({\mathbf x}_1)$ and $p_2({\mathbf x}_2)$ in front
of the detectors. We will end up taking these apertures to be of
infinite width, but initially we leave them in, for reasons to be
explained below. A lens of focal length $f_d$ is placed one focal
length in front of each detector. The distances from the Fourier
plane of the main part of the apparatus to the aperture and from
the aperture to the lens are $d_1$ and $d_2$. In order to erase
path information for the photons reaching each detector, a
polarizer at $45^\circ$ to the polarization directions of both
incoming beams is placed in each path. The two polarizers are
oriented orthogonal to each other.

\begin{figure}
\centering
\includegraphics[totalheight=2in]{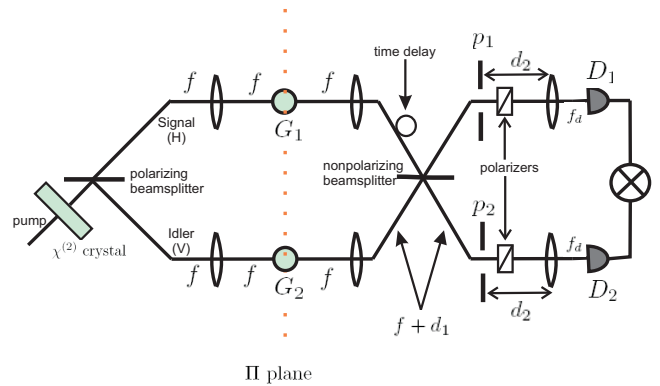}
\caption{\textit{(Color online) Schematic view of aberration-cancellation setup. (Distances and angles not necessarily
drawn in correct proportions.)
The horizontally polarized signal travels in the upper branch and
experiences modulation $G_1$, while the vertically polarized idler experiences modulation $G_2$ in the lower branch. $G_1$ and $G_2$ are both located in the plane $\Pi$, halfway between the lenses of focal length $f$. The beam splitter mixes the beams
before they reach the detectors $D_1$ and $D_2$, which are connected by a coincidence circuit.}}
\end{figure}

The full transfer function for each branch is \cite{bonato2}
\begin{equation} H_{j\alpha}({\mathbf x_\alpha},{\mathbf q_j},\omega )=
G_j({f\over k}{\mathbf q_j})H_{D_\alpha}({\mathbf
x_\alpha},{\mathbf q_j},\omega ),\label{transfer}\end{equation}
where the transfer function of the detection stage is
\begin{eqnarray} H_{D_\alpha}({\mathbf x_\alpha},{\mathbf
q_j},\omega )&=&e^{ik(d_1+d_2+f_D)} e^{-{{ik}\over {2f_D}}\left(
{{d_2}\over
{f_D}}-1\right) {\mathbf x_\alpha}^2}\nonumber \\
& & \quad \cdot e^{-{{id_1}\over {2k}}{\mathbf q_j}^2} \tilde
{\cal P}_\alpha \left({k\over {f_D}}{\mathbf x_\alpha}-{\mathbf
q_j}\right) .\label{detectortrans}\end{eqnarray} $\tilde {\cal
P}_\alpha$ is the Fourier transform of the aperture function,
\begin{equation} \tilde {\cal P}_\alpha\left({k\over
{f_D}}{\mathbf x_\alpha}-{\mathbf q_j}\right)=\int d^2x^\prime
p_\alpha ({\mathbf x}^\prime ) e^{-i\left({k\over {f_D}}{\mathbf
x_\alpha}-{\mathbf q_j}\right)\cdot {\mathbf x^\prime}}
,\end{equation} with $\alpha=\{ 1,2\}$ labelling the detector and
$j=\{ s,i\} $ labelling the signal or idler branch. In these
expressions, $k$ is the longitudinal wavenumber, $k=\sqrt{(\omega
/c)^2-q^2}\approx {\omega\over c}$ for $|{\mathbf q}|<<k$.

The nonpolarizing beam splitter mixes the incident beams, so each detector sees
a superposition of the signal and idler beams. The
positive-frequency part of the field entering detector $\alpha$ is
given by
\begin{eqnarray} E_\alpha^{(+)} ({\mathbf x}_\alpha,t_\alpha ) &=& \int dq d\omega
e^{-i\omega t_\alpha }\left[ H_{s\alpha}({\mathbf x_\alpha
},{\mathbf q_s},\omega )\hat a_s ({\mathbf q}_s,\omega ) \right. \nonumber \\
& & \left. \quad + H_{i\alpha }({\mathbf x_\alpha},{\mathbf
q_i},\omega )\hat a_i ({\mathbf q}_i,\omega )\right]
.\end{eqnarray} Using this field, we can compute the amplitude for
coincidence detection:
\begin{eqnarray} & & A({\mathbf x}_1,{\mathbf x}_2,t_1,t_2) = \langle
0|E_1^{(+)}({\mathbf x}_1,t_1) E_2^{(+)}({\mathbf x}_2,t_2) |\Psi \rangle \nonumber\\
& & \; = \int d^2q\; d\nu \; \Phi ({\mathbf q},\nu ) \\
& & \; \times \left[ e^{-i(\Omega_o+\nu )t_1} e^{-i(\Omega_o-\nu
)t_2} H_{s1} ({\mathbf x}_1,{\mathbf q},\nu ) H_{i2}
({\mathbf x}_2,-{\mathbf q},-\nu ) \right. \nonumber\\
& & \; + \left. e^{-i(\Omega_o-\nu )t_1} e^{-i(\Omega_o+\nu )t_2}
H_{i1} ({\mathbf x}_1,-{\mathbf q},-\nu ) H_{s2} ({\mathbf
x}_2,{\mathbf q},\nu ) \right] ,\nonumber \end{eqnarray} where
$H_{j\alpha} ({\mathbf x}_\alpha ,{\mathbf q}_j,\Omega_0\pm \nu )$
have been abbreviated by $H_{j\alpha} ({\mathbf x}_\alpha
,{\mathbf q}_j,\pm \nu )$.

The coincidence rate as a function of time delay $\tau$ is \begin{equation} R(\tau ) \; =\; \int d^2x_1 d^2x_2dt_1dt_2 |A({\mathbf x}_1,{\mathbf x}_2,t_1,t_2)|^2
.\end{equation} As was shown in \cite{rubin}, $R(\tau )$ will
generically be of the form
\begin{equation} R(\tau )=R_0\left[
1-\Lambda \left( 1-{{2\tau}\over {DL}}\right) W(\tau )\right]
\label{coincidence}.
\end{equation} where $\Lambda (x)$ is the triangular function:
\begin {equation}
\Lambda (x) =
\bigg \{
\begin{array}{l}
1 - |x|, \qquad |x| \leq 1 \\
0, \qquad \qquad |x| > 1
\end{array} \end{equation}
The $\tau$-independent background term $R_0$ and $\tau$-dependent
modulation term $W(\tau )$ were calculated in \cite{bonato2} to be:
\begin{eqnarray}R_0&=& \int d^2q d^2q^\prime sinc [ ML {\mathbf e}_2\cdot
({\mathbf q}-{\mathbf q^\prime} ) ]\nonumber \\
& & \times \; G_1^\ast \left( {{f{\mathbf q} }\over
k}\right)G_2^\ast \left( -{{f{\mathbf q} }\over k}\right)G_1
\left( {{f{\mathbf q^\prime} }\over k}\right)G_2 \left(
-{{f{\mathbf q^\prime} }\over k}\right)\nonumber \\ & & \times \;
\tilde {\cal P}_1({\mathbf q}-{\mathbf q^\prime} )\tilde {\cal
P}_2 (-{\mathbf q}+{\mathbf q^\prime} ) \nonumber
\\ & & \times \; e^{-{{iML}\over 2}{\mathbf e_2}\cdot ({\mathbf q}-{\mathbf q^\prime} )}
e^{{{2id_1\over {k_{pump}}}}({\mathbf q}^2-{\mathbf q^{\prime 2}})} \label{background1}\\
W(\tau )&=&  {1\over {R_0}}\int d^2q d^2q^\prime sinc \left[ ML
e_2\cdot ({\mathbf q}+{\mathbf q^\prime} ) \Lambda \left(
1-{{2\tau}\over {DL}}\right)
\right] \nonumber \\
& & \times \; G_1^\ast \left( {{f{\mathbf q} }\over
k}\right)G_2^\ast \left( -{{f{\mathbf q} }\over k}\right)G_1
\left( {{f{\mathbf q^\prime} }\over k}\right)G_2
\left( -{{f{\mathbf q^\prime} }\over k}\right)\nonumber \\
& & \times \; \tilde {\cal P}_1({\mathbf q}+{\mathbf q^\prime}
)\tilde {\cal P}_2 (-{\mathbf q}-{\mathbf q^\prime} )\nonumber \\
& & \times \; e^{-{{iM}\over D}\tau {\mathbf e_2}\cdot ({\mathbf
{\mathbf q}}-{\mathbf {\mathbf q^\prime}} )} e^{{{2id_1\over
{k_{pump}}}}({\mathbf q}^2-{\mathbf q^{\prime 2}})}
.\label{modulation1}
\end{eqnarray}

Now let the apertures be large, so that the $\tilde {\cal P}_j$
become delta functions, reducing equations (\ref{background1}) and
\ref{modulation1} to:
\begin{eqnarray} R_0&=& \int d^2q \left| G_1 \left( {{f{\mathbf q} }\over
k}\right) G_2 \left( -{{f{\mathbf q} }\over
k}\right) \right|^2 \label{tau}\label{background2}\\
W(\tau )&=& {1\over {R_0}}\int d^2q e^{-{{2iM\tau}\over D}{\mathbf
e_2}\cdot {\mathbf q}}G_1^\ast \left( {{f{\mathbf q}}\over
k}\right)G_1 \left( -{{f{\mathbf q} }\over
k}\right)\nonumber \\
&  &\qquad \times\; G_2^\ast \left( -{{f{\mathbf q} }\over
k}\right)G_2 \left( {{f{\mathbf q} }\over
k}\right)\label{modulation2} .\end{eqnarray}

Suppose that $G_j({\mathbf x})=t_j({\mathbf x})e^{i\phi_j({\mathbf
x})}$, where $t_j$ is real and the effects of aberrations are
contained in the phase factor $\phi_j$. Disregarding the
background term for the moment, we see from the presence in
equation (\ref{modulation2}) of the factors
\begin{eqnarray} & & G_1^\ast \left( {{f{\mathbf q} }\over
k}\right)G_1 \left( -{{f{\mathbf q} }\over k}\right) \nonumber \\
& & \quad = t_1^\ast\left( {{f{\mathbf q} }\over
k}\right)t_1\left( -{{f{\mathbf q} }\over k}\right)
e^{-i\left[\phi_1\left(  {{f{\mathbf q} }\over
k}\right)-\phi_1\left(-{{f{\mathbf q} }\over
k}\right)\right]}\end{eqnarray} that even order aberration terms
arising from sample 1 cancel from the modulation term. The even
order aberrations from sample 2 cancel similarly. This is the even
order cancellation effect of references \cite{bonato1} and
\cite{bonato2}.

It should be remarked that the setup of figure 1 may be simplified
by removing the lenses immediately in front of the detectors. We
have left both the lenses and the apertures in the setup because
together they lead to the presence of the Fourier transformed
aperture functions $\tilde {\cal P}_j$ in equations
(\ref{background1}) and (\ref{modulation1}); the delta functions that
arise from the $\tilde {\cal P}_j$ when the apertures become large
will serve as convenient bookkeeping devices in the following
sections as we trace various terms back to their origins. If we
choose to simplify the apparatus and remove the lenses, then
equation (\ref{detectortrans}) will be replaced by
\begin{eqnarray}H_{D_\alpha}({\mathbf x}_\alpha ,{\mathbf q}_j, \omega )& =&
e^{ik(d_1+d)} e^{{-id_1{\mathbf q}^2_j}\over {2k}}\\
& & \; \times \; \int p({\mathbf x^\prime} ) e^{{{ik}\over {2d}}
({\mathbf x^\prime} -{\mathbf x}_\alpha )^2}e^{i{\mathbf q}\cdot
{\mathbf x^\prime}} d^2x^\prime ,\nonumber \end{eqnarray} where
$d$ is the total aperture-to-detector distance, with corresponding
changes in equations (\ref{background1}) and (\ref{modulation1}).
However, in the large-aperture limit this does not affect the
coincidence rate, which will still be given by expressions
(\ref{coincidence}), (\ref{background2}), and (\ref{modulation2}).

\section{All-order cancellation}

\subsection{Aberration cancellation to all orders}
Now, consider the background term $R_0$ in equation
(\ref{background2}). It depends on $G_1$ and $G_2$ only through the
squared modulus of each. Thus any phase changes introduced by
$G_1$ or $G_2$ cancel completely; in particular, the background
term $R_0$ exhibits cancellation of aberrations of {\it all}
orders, not just even orders. In the current situation, this $R_0$
is of no importance, simply being a constant and having no effect
on the $\tau$-dependence of the correlation. However, the fact
that all orders of aberration can be cancelled in the background
term raises the question as to whether it can be arranged for this
to happen in the modulation term as well.

It turns out that the answer to this question is positive: it is
possible to use this apparatus to cancel {\it all} aberrations
induced by a thin sample, of both even and odd orders. The means
for doing so is evident from examining equation (\ref{modulation2}).
Suppose that $G_1({\mathbf x})=G_2({\mathbf x})$, as shown
schematically in figure 2. This can can happen in one of two ways:
either two identical samples may be placed in the two
arms, or it may be arranged so that the two beams both pass
through the same sample;
in either case it is necessary for the sample to act in the same
manner on both polarization states. The second possibility will
usually be of more practical interest, since identical samples
will often not be available. For $G_1=G_2$, equations
(\ref{background1}) and (\ref{modulation1}) give {\small
\begin{eqnarray} R_0&=& \int d^2q
\left| G_1 \left( {{f{\mathbf q} }\over k}\right) G_1 \left( -{{f{\mathbf q} }\over
k}\right) \right|^2 \label{Rx}\\
W(\tau )&=& {1\over {R_0}}\int d^2q  e^{-{{2iM\tau \hat {\mathbf e_2}\cdot
{\mathbf q}}\over D}}\left| G_1 \left( {{f{\mathbf q} }\over k}\right) G_1 \left(
-{{f{\mathbf q} }\over k}\right) \right|^2 \label{Wx}
\end{eqnarray}} Setting $G_1({\mathbf x})=t({\mathbf x})e^{i\phi
({\mathbf x})}, $ we see that all phases now cancel from the
$\tau$-modulated term $W$. Thus, {\it all aberrations induced by
the sample, of any order, will completely cancel from the
coincidence rate}.

\begin{figure}
\centering
\includegraphics[totalheight=1.6in, width=3in]{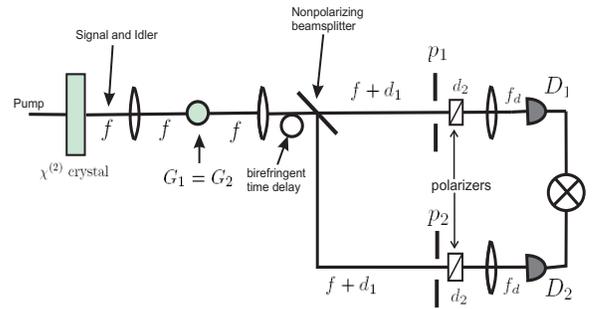}
\caption{\textit{(Color online) Schematic view of apparatus in figure 1, with $G_1$ set equal to $G_2$. (Distances and angles not necessarily
drawn in correct proportions.) Here $G_1$ and $G_2$ are being produced by a single object. The signal and idler are collinear. It is also possible for $G_2$ and $G_2$ to be produced by two identical but spatially separate objects interacting with noncollinear signal and idler.}}
\end{figure}

\subsection{Condition for All-Order Cancellation}

Up to this point, we have assumed that the objects providing the
modulation were located in the plane labelled $\Pi$ in figure 1.
Now we consider what happens if the modulation objects (the
samples) are moved out of the $\Pi$-plane by some distance $z\ne
0$. Consider a single arm of the apparatus, as shown in figure 3.
We will take the distance $z$ from $\Pi$ to be positive if the
sample is moved toward the source, and negative if moved toward
the detector. Now, the impulse response functions for the first
and second lens respectively in each branch of the system will be:
\begin{eqnarray} h_1({\mathbf \xi},{\mathbf y})&=& {1\over {i\lambda f}}\cdot {1\over {i\lambda (f-z)}}\int e^{{{ik}\over 2} \left(
{{{\mathbf y}^2}\over {f-z}}+{{{\mathbf \xi}^2}\over f}\right)} \nonumber \\
& & \times \;e^{ {{ik}\over 2} \left( {1\over
{f-z}}\right){\mathbf x^{\prime 2}}} e^{-ik{\mathbf x^\prime}
\cdot \left( {{\mathbf y}\over {f-z}}+{{\mathbf \xi}\over
f}\right)}\; d^2x^\prime \\
h_2({\mathbf y},{\mathbf x})&=& {1\over {i\lambda f}}\cdot {1\over
{i\lambda (f+z)}}\int e^{{{ik}\over 2} \left( {{{\mathbf
y}^2}\over {f+z}}+{{{\mathbf x}^2}\over f}\right)} \nonumber \\
& & \times \; e^{ {{ik}\over 2} \left( {1\over
{f+z}}\right){\mathbf x}^{\prime \prime 2}}e^{-ik{\mathbf
x^{\prime \prime}} \cdot \left( {{\mathbf y}\over {f+z}}+{{\mathbf
x}\over
f}\right)}\; d^2x^{\prime \prime} 
\end{eqnarray}
${\mathbf y}$, ${\mathbf x^\prime}$, ${\mathbf x^{\prime \prime
}}$, and ${\mathbf \xi}$ are the transverse distances at the
points shown in figure 2. The integrals can be carried out, giving
us the result that:
\begin{eqnarray} h_1({\mathbf \xi},{\mathbf y})&= & {1\over {i\lambda f}}e^{{{ik}\over {2f}}\left[
{z\over f}{\mathbf \xi}^2-2{\mathbf \xi}\cdot {\mathbf y}\right]}\\
h_2({\mathbf y},{\mathbf x})&=  &{1\over {i\lambda
f}}e^{-{{ik}\over {2f}}\left[ {z\over f}{\mathbf x}^2+2{\mathbf
x}\cdot {\mathbf y}\right]} \; =\; -h_1^\ast (-{\mathbf
x},{\mathbf y}).\end{eqnarray} So the impulse response for one
branch of the apparatus from source to Fourier plane (not
including the detection stage) is
\begin{eqnarray} h_j^\prime ({\mathbf \xi} ,{\mathbf x}) &=& \int h_1({\mathbf \xi} ,{\mathbf y})G_j({\mathbf y})h_2({\mathbf y},{\mathbf x})d^2y \\
&=& {{e^{{{ik}\over {2f^2}} z ({\mathbf \xi}^2-{\mathbf
x}^2)}}\over {(i\lambda f)^2}}\int e^{-{{ik}\over {f}}\left(
{\mathbf \xi} +{\mathbf x}\right)\cdot {\mathbf y}}G_j({\mathbf
y})d^2y .\end{eqnarray} Fourier transforming to find the transfer
function leads to:
\begin{eqnarray} H_j^\prime ({\mathbf x},{\mathbf q},\omega ) &=& \int h({\mathbf \xi}
,{\mathbf x})\; e^{i{\mathbf q}\cdot{\mathbf \xi} } d^2\xi \\
&=& {1\over {(i\lambda f)^2}}\int d^2y \; G_j({\mathbf y})\;
e^{-{{ik}\over f}{\mathbf x}\cdot {\mathbf y}} e^{-{{ik}\over
{2f^2}}z{\mathbf x}^2}\nonumber \\
& & \qquad \times  \int d^2 {\mathbf \xi} \; e^{{{ikz}\over {2f^2}}
{\mathbf \xi}^2}e^{ i{\mathbf \xi}
\left( {\mathbf q}-{{k{\mathbf x}}\over f}\right)}\\
&=& -{1\over {\lambda z}}e^{-i{\mathbf q}\cdot {\mathbf x}}\int d^2y \; G_j\left( {\mathbf y}+{{f{\mathbf q}}\over k} \right) \nonumber \\
& & \qquad \times \; e^{-{{ik}\over f} {\mathbf x}\cdot {\mathbf
y}}e^{-{{ ik}\over {2z}}{\mathbf y}^2} e^{-{{ikz}\over
{2f^2}}{\mathbf x}^2}.\end{eqnarray}

\begin{figure}
\centering
\includegraphics[totalheight=2in]{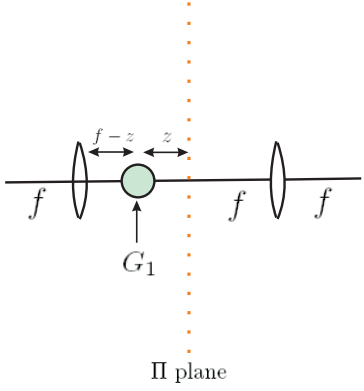}
\caption{\textit{(Color online) Blown up version of a portion of one branch from apparatus of figure 1 (or figure 2), with the object moved a
distance $z$ out of the central plane, $\Pi$.}}
\end{figure}

Previously, for $z=0$, this transfer function was simply given by
\begin{equation} H_j^\prime ({\mathbf x},{\mathbf q},\omega )=(constants)\cdot G_j\left({{f{\mathbf q}}\over k}\right) e^{-i{\mathbf q}\cdot
{\mathbf x}}.\end{equation} Therefore, for $z\ne 0$, we must make
the replacement (up to overall constants)
\begin{eqnarray} G\left({{f{\mathbf q}}\over k}\right)&\longrightarrow &\int
d^2y \; G\left({\mathbf y}+{{f{\mathbf q}}\over k}\right)\\
& & \qquad \times \;e^{-{{ik}\over f}{\mathbf x}\cdot {\mathbf
y}}e^{-{{ikz}\over {2f^2}}{\mathbf x}^2}\left( {1\over
z}e^{-{{ik}\over {2z}}{\mathbf y}^2}\right)\nonumber
\end{eqnarray} in all previous results, and equation
(\ref{transfer}) now involves an integral instead of a simple
product. (For $z=0$, the factors in the last set of parentheses
become proportional to $\delta^{(2)} ({\mathbf y})$, leading back
to the previous results.) In particular, in equations (\ref{Rx})
and (\ref{Wx}), the factor $\left| G_1\left({{f{\mathbf q}}\over
k}\right)\right|^2$ becomes
\begin{equation} \int d^2y \; d^2y^\prime \; G_1\left({\mathbf y}+{f\over
{f_D}}{\mathbf x}\right)G_1^\ast\left({\mathbf y}^\prime +{f\over {f_D}}{\mathbf x}\right)
e^{-{{ik}\over {2z}}({\mathbf y}^2-{\mathbf y^{\prime 2}})}.\label{outofplane}
\end{equation} Clearly, the phase of $G_1$ no longer cancels out
of this expression since nothing forces ${\mathbf y}$ to equal ${\mathbf y^\prime}$.
The arguments of the two factors of $G_1$ are now unrelated, so
that aberration cancellation no longer occurs.

So any cancellation that occurs can hold exactly {\it only} for
phases arising in the $\Pi$-plane of the Fourier transform
system. The cancellation is approximate in the vicinity of this
plane. For
samples of finite thickness, the degree of approximate
cancellation will diminish as the thickness increases.

Defining ${\mathbf\epsilon}={\mathbf y}- {\mathbf y^\prime}$, the
exponential term in equation (\ref{outofplane}) becomes
\begin{equation} e^{-{{ik}\over {2z}}\left( 2{\mathbf
\epsilon}\cdot {\mathbf y}-{\mathbf \epsilon }^2 \right)}.
\end{equation} Assuming that $G_1^\ast \left( {\mathbf y}-\epsilon
+{f\over {f_d}}{\mathbf x}\right) $ is slowly varying in ${\mathbf
\epsilon}$ compared to the variation of the exponential, we may
obtain an estimate of the distance $z$ over which the sample may
be moved out of the plane while still maintaining a high degree of
abberation cancellation.  The aberration cancels when ${\mathbf
\epsilon}=0$, so we may use the maximum size of ${\mathbf
\epsilon}$ as a measure of the degree of failure of the aberration
cancellation. As $z \to 0$, the rapid oscillations of the
exponential term cause the integral of equation (\ref{outofplane})
to go to zero, unless $ k\left| 2{\mathbf \epsilon}\cdot {\mathbf
y}-{\mathbf \epsilon }^2\right| $ also goes to zero at least as
fast as $|z|$. So, we must have \begin{equation}\left| 2{\mathbf
\epsilon} \cdot {\mathbf y}-{\mathbf \epsilon}^2\right| \lesssim
\left| {z\over k}\right| \sim |z\lambda |.\end{equation} From
this, we have \begin{equation} |z|\sim {{{\mathbf \epsilon}_M \;
|{\mathbf y}|}\over \lambda},\label{estimate}\end{equation} where
${\mathbf \epsilon}_M$ is the maximum value of ${\mathbf
\epsilon}$. Let $r_s$ be the maximum illuminated radius of the
sample. Then, by requiring that $|{\mathbf \epsilon}_M|<<r_s$, we
have the estimate that \begin{equation}|z|<<{{r_s^2}\over
\lambda}.\end{equation} This is essentially a limit on how far
from stationarity we may be and still safely apply a
stationary-phase approximation. Actually, we may make this limit a
bit more precise. Since two sample points ${\mathbf y}$ and
${\mathbf y^\prime}$ inside the Airy disk of the lens can not be
distinguished from each other, we may require that $|{\mathbf
\epsilon}_M|\sim R_{airy}$, where
\begin{equation} R_{airy} ={{1.22f\lambda}\over a}\end{equation}
is the radius of the Airy disk. By substituting this into equation
(\ref{estimate}), we can thus conclude that, at most, the order of
magnitude of $|z|$ may be given by
\begin{equation} |z|\lesssim {{fr_s}\over a}.\end{equation} Taking
for example the values $r_s\sim 10^{-4}m$, $a\sim 1 \; cm$, $f\sim
10\; cm$, and $\lambda \sim 10^{-7}\; m$, this gives us an upper
limit of about $1\; mm$.

\subsection{Comparison with Dispersion Cancellation}

The idea of aberration cancellation via entangled-photon
interferometry arose in analogy to the similar dispersion
cancellation effect \cite{franson}, \cite{steinberg}. It is known
that even-order and odd-order dispersion effects may be separated
so that either even-order terms or odd-order terms may be
cancelled \cite{minaeva1}, but that it is impossible to
simultaneously cancel both sets of terms together. Thus, it is a
surprise that in the case of aberrations such a simultaneous
cancellation should be possible.

The fact that aberration cancellation only occurs in a single
plane sheds some light on the difference between aberration
cancellation and dispersion cancellation. Aberrations are caused
by phase differences between different points in a plane {\it
transverse to} the propagation direction of the light, while
dispersion comes about as a result of phase differences
accumulating {\it along} the propagation direction. We have
managed to cancel all orders of aberration produced by {\it a
single transverse plane}. But since dispersive effects accumulate longitudinally,
we cannot arrange their cancellation in all of the infinite number of transverse planes
the photon travels through; thus, although even-order and odd-order dispersion
may each occur separately, simultaneous all-order dispersion
cancellation will not occur.

A more physical explanation can be given for the inability in
principle to cancel all orders of dispersion. Suppose that the
index of refraction is expanded about some frequency $\omega_0$,
\begin{equation} n(\omega )=n_0+n_1(\omega-\omega_0)+n_2(\omega-\omega_0)^2+\dots \end{equation}
The phase and group velocities are \begin{eqnarray}
v_p&=& {c\over {n(\omega )}} \\
v_g&=& \left( {{dk}\over {d\omega}} \right)^{-1}
\; =\; c\left[ n(\omega ) +\omega  {{dn(\omega )}\over {d\omega }}\right]^{-1} \nonumber\\
&=& c\left[
n_0+2n_1(\omega-\omega_0)+3n_2(\omega-\omega_0)^2+\dots
\right]^{-1}.\end{eqnarray} If both the odd-order and even-order
dispersion coefficients vanish simultaneously (including the
zeroth-order term), then $n(\omega )$ and $ {{dn}\over {d\omega}}
$ both vanish. In consequence, the phase and group velocities both
diverge. This is in contradiction to special relativity, which
requires a finite group velocity. In contrast, no similar obstacle
exists to prevent the spatially distributed phase shift $\phi
({\mathbf x})$ from vanishing, so there is no fundamental
principle preventing all-order aberration cancellation.


One further point to note is that the dispersive and aberrative
cases considered here are not entirely analogous, in the sense
that one is not simply obtained from the other by interchanging
time and space. In the aberration case, the phase is a function of
the transverse position ${\mathbf x}$ in the physical coordinate
space. In contrast, for the dispersive case the phase is due to a
{\it frequency}-dependent index of refraction; i.e. the source of
the effect is in the {\it Fourier transform space}, not in the
(temporal) coordinate space. However, in both cases the
cancellation occurs in the Fourier space. Thus, for aberration
cancellation an optical Fourier transform system is required to
move from the coordinate space (where the source of aberration is)
to the Fourier space (where the cancellation occurs). For the
dispersive case, the source of the dispersion already operates in
the Fourier space so it is not necessary to introduce an extra
Fourier transform via the optical system.

\section{Physical Interpretation}

We now wish to develop a better understanding of how aberration
cancellation occurs in the polarization-based coincidence
interferometer that we are using to illustrate this effect.

Let ${\mathbf q}$ and ${\mathbf q^\prime} $ be the ingoing and
outgoing momenta in the upper branch at the beam splitter. The
ingoing and outgoing momenta for the lower branch will be
$-{\mathbf q}$ and $-{\mathbf q^\prime}$, as in figures 4 and 5
below.

Note first of all that the coincidence detection amplitude in
transverse momentum space may be written in the form $A({\mathbf
q})=A_r({\mathbf q})+A_t({\mathbf q})$, where $A_t$
represents the amplitude for both photons to be transmitted at the
beam splitter and $A_r$ is the amplitude for both to be reflected.
The counting rate involves the integrated and squared amplitude;
if the momenta ${\mathbf q}$ and ${\mathbf q^\prime}$ were
independent variables, we could write this as
\begin{equation}\left|\int A({\mathbf q})d^2q\right|^2=\int
A({\mathbf q})A^\ast ({\mathbf q^\prime} )d^2q d^2q^\prime
,\end{equation} which has terms $A_r({\mathbf q}) A_t({\mathbf
q^\prime} )^\ast +A_t ({\mathbf q})A_r^\ast ({\mathbf q^\prime} )$
involving interference between reflection and transmission (see
figure 4), as well as non-interference terms $A_r({\mathbf q})
A_r({\mathbf q^\prime})^\ast +A_t ({\mathbf q})A_t^\ast ({\mathbf
q^\prime} )$ (figure 5). However, ${\mathbf q}$ and ${\mathbf
q^\prime} $ are not independent variables; momentum conservation
and the fact that the photons are produced from downconversion
together force the requirement ${\mathbf q^\prime} =\pm {\mathbf
q}$. These constraints are explicitly enforced in the current
context by the factors of $\tilde {\cal P}_j$ in equations
(\ref{background1}) and (\ref{modulation1}), which become delta
functions in the large aperture limit. The delta functions sew together the
amplitudes $A_r$ and $A_t$ as shown in the figures.

Suppose again that $G_j({\mathbf x})=t_j({\mathbf
x})e^{i\phi_j({\mathbf x})}$. Since we are unconcerned with
effects related to amplitude modulation we henceforth set
$t_j({\mathbf x})=1$. Examining equations (\ref{background1}) and
(\ref{modulation1}), we then note that even-order and odd-order
aberration cancellation arise from different sources. Even-order
cancellation arises from the combination of the following
ingredients:

\vskip 10pt A1. The Fourier transforming property of the lens in
the focal plane. This converts the transverse momentum entanglement into
spatial entanglement in the $\Pi$-plane.

A2. The condition ${\mathbf q}=-{\mathbf q^\prime}$ satisfied by
the non-background half of the terms (those that comprise $W$).
These terms arise from the interference part of the squared
amplitude, as in figure 4.

A3. The $G_j\left({{f{\mathbf q}}\over
k}\right)G_j^\ast\left({{f{\mathbf q^\prime} }\over k}\right)$
structure that arises from taking the absolute square of the
amplitude to find counting rates in quantum mechanics $(j=1,2)$.
Combined with the momentum constraint of A2, this becomes
$G_j\left({{f{\mathbf q}}\over k}\right)G_j^\ast\left(
-{{f{\mathbf q} }\over k}\right)=e^{i\left[ \phi_j({\mathbf
q})-\phi_j(-{\mathbf q})\right]}.$ \vskip 10pt

\begin{figure}
\centering
\includegraphics[totalheight=2in]{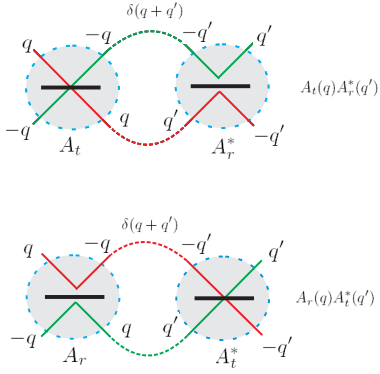}
\caption{\textit{(Color online) Schematic representation of interference terms. In the squared amplitude
$\int dq\; dq^\prime A(q)A^\ast (q^\prime )$, the part of the
amplitude in which both photons undergo reflection at the
beam splitter ($A_r$) interferes with the portion in which both
photons are transmitted at the beam splitter $(A_t)$. For these
terms, $q=-q^\prime$, due to the delta function that
connects the amplitudes.}}
\end{figure}

\begin{figure}
\centering
\includegraphics[totalheight=2in]{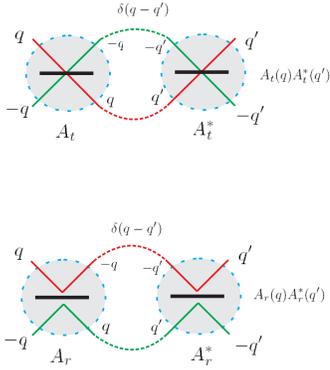}
\caption{\textit{(Color online) Schematic representation of noninterference terms. In the top part of the
figure the transmission portion of the amplitude $A_t$ interacts
only with itself, while in the bottom part the same is true of the
reflection amplitude $A_r$. For these terms, $q=q^\prime$}}
\end{figure}

In contrast, odd-order cancellation occurs when the following
combination of ingredients is present:

\vskip 10pt B1. The Fourier transforming action of the lens, as in
A1.

B2. For every photon of transverse momentum ${\mathbf q}$ there is
a photon of $-{\mathbf q}$ present due to downconversion.

B3. $G_1=G_2$, so that the product $G_1\left({{fq}\over k}\right)
G_2\left(-{{f{\mathbf q}}\over k}\right) $ becomes
$G_1\left({{f{\mathbf q}}\over k}\right) G_1 \left(-{{f{\mathbf
q}}\over k}\right) =e^{i\left[ \phi_1({\mathbf
q})+\phi_1(-{\mathbf q})\right]}.$ (Note that the cancellation is
taking place between different terms of equation (\ref{modulation2})
than were involved in the cancellation of A3.)\vskip 10pt

In order to have all-order cancellation, there are two
possibilities. Either both of the above sets of conditions may be
satisfied simultaneously, or else a third set of conditions may be
satisfied:

C1. Same as A1 and B1.

C2. The condition ${\mathbf q}={\mathbf q^\prime} $ must be
satisfied, as in the background term $R_0$. This occurs in the
noninterference terms of figure 5.

C3. Similar to A3, the $G_j\left({{f{\mathbf q}}\over
k}\right)G_j^\ast\left({{f{\mathbf q^\prime} }\over k}\right)$
structure arises from the quantum mechanical absolute squaring of
the amplitude. But now, coupled with C2, we have
$G_j\left({{f{\mathbf q}}\over k}\right)G_j^\ast\left( {{f{\mathbf
q} }\over k}\right)=e^{i\left[ \phi_j({\mathbf q})-\phi_j({\mathbf
q})\right]} =1$, giving cancellation of all orders. \vskip 10pt


In A3 and C3 the phase from a single arm of the interferometer
cancels with itself, whereas B3 is a cancellation between the two
different (but identical in this case) arms. Cases A and B both
involve interference between the amplitudes $A_r$ and $A_t$ (shown
schematically in figure 4), while case C comes from the
non-interference terms of figure 5, and so will occur even if only
one of the two amplitudes $A_r$ and $A_t$ is present.

\section{Conclusions}

To summarize the main results of this paper, for the apparatus
of figure 1 we have found that:

$\bullet$ Even-order aberrations induced by the samples $G_1$ and $G_2$ cancel.

$\bullet$ If the two beams overlap so that $G_1=G_2$, then {\it all}
orders of aberration cancel.

$\bullet$ These cancellations only occur if $G_1$ and $G_2$ are
confined to the $z=0$ plane.

These results open up the possibility of using quantum
interferometry to eliminate the effects of sample-induced
aberration in experiments using temporal correlation-based methods
such as dynamical light scattering or fluorescence correlation
spectroscopy.

Through the continued study of aberration-cancellation and
dispersion-cancellation, it is hoped that a better understanding
of the effects of objects or materials placed in an optical
system, and better methods of controlling those effects, will
gradually emerge. The results reported here are one more step
along that path.

The effects described in this paper make essential use of the
spatial entanglement (or equivalently the transverse momentum
entanglement) between the photons in the downconversion pair. In
contrast, the frequency entanglement played no essential role.
Similarly, the anticorrelation of the polarizations was used
primarily to control the paths of the photons and then to erase
the path information; but these functions could be accomplished by
other means. So only the spatial entanglement was essential.
On the other hand, it is the frequency entanglement that is essential
for dispersion cancellation. A question for future investigation is
whether use of the simultaneous entanglement of frequency,
momentum, and polarization variables (so-called {\it
hyperentanglement}) may allow control over further optical effects
of materials.

\vskip 25pt
\begin{acknowledgments}
This work was supported by a U. S. Army Research Office (ARO)
Multidisciplinary University Research Initiative (MURI) Grant; by
the Bernard M. Gordon Center for Subsurface Sensing and Imaging
Systems (CenSSIS), an NSF Engineering Research Center; by the
Intelligence Advanced Research Projects Activity (IARPA) and ARO
through Grant No. W911NF-07-1-0629.
\end{acknowledgments}

\end{document}